# SPATIAL AND TEMPORAL NOISE SPECTRA OF SPATIALLY EXTENDED SYSTEMS WITH ORDER-DISORDER PHASE TRANSITIONS

K.Staliunas
Physikalisch Technische Bundesanstalt, 38116 Braunschweig, Germany
tel.: +49-531-5924482, Fax: +49-531-5924423, E-mail: Kestutis.Staliunas@PTB.DE

**Abstract**

The noise power spectra of spatially extended dynamical systems are investigated, using as a model the Complex Ginzburg-Landau equation with a stochastic term. Analytical and numerical investigations show that the spatial spectra of the ordered state are similar to Bose-Einstein distribution, showing $1/k^2$ asymptotics in the long wavelength limit. The temporal noise spectra of the ordered state obtained are of $1/f^a$ form, where $a = 2 - D/2$ with $D$ the spatial dimension of the system.

## 1. Introduction

In this article the temporal and spatial noise spectra of the spatially extended nonlinear systems with order-disorder transitions are investigated. Order-disorder transitions are universally described in the lowest order by a complex Ginzburg - Landau equation (CGLE) with stochastic Langevin forces. The universality of the CGLE model for describing order - disorder transitions is discussed in Section 2.

The analytical and numerical study of the spatial and temporal spectra of the ordered (homogeneous, nontrivial) solution of the CGLE given here leads to several surprising results:

1) The spatial noise spectra of the CGLE results in a $1/k^2$ - form, where k is the spatial wavenumber of the perturbation of the homogeneous state. Thus, in the long wavelength limit ($k \to 0$) the statistical distributions coincide with the Bose-Einstein distribution, derived originally for systems in thermal equilibrium [S.N.Bose, 1924]. Consequently the assumption of thermal equilibrium seems to be not necessary in order to obtain statistical Bose-Einstein distributions, which are apparently related with order-disorder transitions in equilibrium as well as in nonequilibrium systems.

2) The temporal noise spectra of the CGLE results in a $1/f^a$ - form, where $a = 2 - D/2$ with $D$ the spatial dimension of the system. Thus 1/f spectra are found to occur very generally and naturally in spatially extended systems with order-disorder transitions. The universality of 1/f-, or flicker noise is an old puzzle of physics [P.Dutta & P.M.Horn, 1981; Sh.M.Kogan, 1985, M.B.Weissman, 1985; G.P.Zhigal'skii, 1997], and a universal explanation of it has never been given. The results reported here might help to elucidate this puzzle.

Spatio-temporal noise spectra are obtained analytically in Section 3 as following from simple linear stability analysis of the ordered solution of the CGLE. Section 4 gives numerical results. Numerical proof of the analytically predicted $1/f^a$ temporal spectra and of $1/k^2$ spatial spectra as extending over many decades of temporal and spatial frequencies is not trivial, especially in cases of larger space dimensions, considering that finite integration grids must be used. The numerical problems have been solved by developing a multiscale numerical technique, as described in Section 4.

These power-law spatial and temporal spectra lead to divergences over the integrals of the spectra. Obviously low frequency divergence of the integral over the temporal spectra leads to development of singularities in the patterns for long times. Such singularities occur for low dimensional patterns: for $a > 1$, consequently for $D < 2$. For high dimensional systems (for $a < 1$, and for $D > 2$) the integral over the temporal spectra diverges for high frequencies causing the singularities at small times. These low- and high frequency divergences are discussed in the conclusions.

## 2. Model for order-disorder phase transitions

As mentioned above order-disorder transitions in spatially extended nonlinear systems can be described in the lowest order by a complex Ginzburg - Landau equation (CGLE) with a stochastic term:

$$\frac{\partial A}{\partial t} = pA - (1+ic)|A|^2 A + (1+ib)\nabla^2 A + \Gamma(\mathbf{r},t) \tag{1}$$

Here $A(\mathbf{r},t)$ is the complex-valued order parameter defined in n-dimensional space $\mathbf{r}$, and evolving with time $t$. $p$ is the control parameter (the order-disorder transition occurs at $p = 0$). The Laplace operator $\nabla^2 A$ represents nonlocality in the system, and $\Gamma(\mathbf{r},t)$ is an additive noise, $\boldsymbol{d}$ - correlated in space and time and of temperature $T$: $\langle \Gamma(\mathbf{r}_1, t_1) \cdot \Gamma^*(\mathbf{r}_2, t_2) \rangle = 2T \cdot \boldsymbol{d}(\mathbf{r}_1 - \mathbf{r}_2) \boldsymbol{d}(t_1 - t_2)$.

Below the transition threshold ($p < 0$) CGLE (1) yields a disordered state: the order parameter $A(\mathbf{r}, t)$ is essentially a noise filtered in space and time, with exponential (thermal) intensity distribution. Above the transition threshold ($p > 0$) (1) yields an ordered, or coherent state (or a condensate) in modulationally stable cases, with the intensity distributed around its mean value $\langle |A|^2 \rangle = p$.

CGLE (1), with complex-valued coefficients, has been derived systematically for many systems showing second order phase transitions in the presence of noise, e.g.: for lasers with spatial degrees of freedom [R.Graham & H.Haken, 1970], where $A(\mathbf{r}, t)$ is proportional to the amplitude of the optical field, and $\Gamma(\mathbf{r}, t)$ corresponds to the vacuum- or thermal fluctuations; for finite temperature superfluids [V.L.Ginzburg & L.D.Landau, 1950; V.L.Ginzburg & L.P.Pitaevskii, 1958], and for finite temperature Bose-Einstein condensates [E.P.Gross, 1961; L.P.Pitaevskii, 1961], where $A(\mathbf{r}, t)$ is the wave-function of the condensate, and $\Gamma(\mathbf{r}, t)$ corresponds to the thermal bath. The CGLE with real-valued coefficients $b = c = 0$ (however, with complex-valued order parameter), has been also systematically derived as the amplitude equation for stripe patterns in nonequilibrium dynamical systems [A.S.Newell & J.A.Whitehead, 1969; L.A.Segel, 1969], where the amplitude and phase of the order parameter corresponds to the amplitude- and phase modulations of the stripe patterns respectively.

The CGLE, with real-valued coefficients, can be written phenomenologically as a normal form, or the minimal equation, describing universally nonequilibrium order-disorder phase transitions [H.Haken, 1979; Y.Kuramoto, 1984; P.Manneville, 1990]: the first two terms $(pA - |A|^2 A)$ approximate in the lowest order a supercritical Hopf bifurcation, a bifurcation bringing the system from a trivial state to a state with phase-invariance of the order parameter $A(\mathbf{r}, t)$. The complex-valued character of the order parameter is important, since every ordered, or coherent state, both in classical or quantum mechanics, is characterized not only by the modulus of the order parameter, but also by its phase. The diffusion term $\nabla^2 A$ describes the simplest possible nonlocality in spatially isotropic and translationally invariant systems.

The *real* Ginzburg-Landau equation was introduced [L.D.Landau & E.M.Lifschitz, 1959] as the normal form for second order phase transitions between two arbitrary spatially extended states and can be derived systematically for many systems, as well as phenomenologically from symmetry considerations. The real Ginzburg-Landau equation does not contain information on the coherence properties of the system. Thus analogously we try to find a simple model for the order-disorder phase transition, a normal form that can be derived systematically for particular systems, as well as phenomenologically from symmetry considerations. The *complex* Ginzburg-Landau equation is just that. It describes as a normal form systems characterized by: 1) supercritical phase transition between disordered and ordered state. 2) phase invariance of the order parameter, 3) isotropy and homogeneity of space.

## 3. Spatio-temporal noise spectra

For analytical treatment it is assumed, that the system is sufficiently far above the order-disorder transition: $p \gg T$. Then the homogeneous component $|A_0| = \sqrt{p}$ is dominating, and one can look for a solution of (1) in the form of a perturbed homogeneous state: $A(\mathbf{r},t) = A_0 + a(\mathbf{r},t)$. We assume below $A_0$ real-valued without the loss of generality. After linearisation of (1) around $A_0$, and diagonalisation, one obtains the linear stochastic equations for perturbations: $b_+ = (a + a^*)/\sqrt{2}$ $b_- = (a - a^*)/\sqrt{2}$ respectively:

$$\frac{\partial b_+}{\partial t} = -2pb_+ + \nabla^2 b_+ + \Gamma_+(\mathbf{r},t) \tag{2.a}$$

$$\frac{\partial b_-}{\partial t} = \nabla^2 b_- + \Gamma_-(\mathbf{r},t) \tag{2.b}$$

Here $\Gamma_\pm = (\Gamma \pm \Gamma^*)/\sqrt{2}$. The components of perturbation $b_+$ and $b_-$ are respectively parallel and perpendicular to the vector of homogeneous state, thus have a sense of amplitude and phase perturbations. The amplitude fluctuations $b_+$ decay with the rate: $\lambda_+ = -2p - k^2$ as (2.a) indicates, where $k$ is the spatial wavenumber of the perturbation. Asymptotically long-lived amplitude perturbations are possible only at the Hopf bifurcation point (in the critical state), but never above or below it. The phase fluctuations $b_-$ decay with a rate $\lambda_- = -k^2$ above the Hopf-bifurcation point as (2.b) shows. This means that the long-wavelength modes decay asymptotically slowly, with a decay rate approaching zero for $k \to 0$, which is a consequence of the phase invariance of the system. The phase, in this way, is in a critical state for all $p > 0$.

From (2) one can calculate spatio-temporal noise spectra, by rewriting (2) in terms of the spatial and temporal Fourier components $b_\pm(\mathbf{r},t) = \int b_\pm(\mathbf{k},w) \exp(iwt - i\mathbf{kr}) dw d\mathbf{k}$:

$$b_+(\mathbf{k},w) = \frac{\Gamma_+(\mathbf{k},w)}{iw + \mathbf{k}^2 + 2p}; \qquad b_-(\mathbf{k},w) = \frac{\Gamma_-(\mathbf{k},w)}{iw + \mathbf{k}^2} \tag{3}$$

The spatio-temporal power spectra are:

$$S_+(\mathbf{k},w) = |b_+(\mathbf{k},w)|^2 = \frac{|\Gamma_+(\mathbf{k},w)|^2}{w^2 + (2p + \mathbf{k}^2)^2}; \qquad S_-(\mathbf{k},w) = |b_-(\mathbf{k},w)|^2 = \frac{|\Gamma_-(\mathbf{k},w)|^2}{w^2 + \mathbf{k}^4} \tag{4}$$

for the amplitude- and phase fluctuations correspondingly. Assuming $\delta$-correlated noise in space and time, $|\Gamma_\pm(\mathbf{k},w)|^2$ are simply proportional to the temperature $T$ of the random force.

**a) spatial power spectra**

The spatial spectra are obtained by integration (4) over all temporal frequencies $w$: $S(\mathbf{k}) = S_+(\mathbf{k}) + S_-(\mathbf{k}) = \int S_+(\mathbf{k},w) dw + \int S_-(\mathbf{k},w) dw$. (The total power spectrum here is the sum of amplitude $S_+(\mathbf{k})$ and phase $S_-(\mathbf{k})$ power spectra, since the spectral components $b_\pm(\mathbf{r},t)$ are mutually uncorrelated, as follows from (3)). The integration is performed for clarity separately for amplitude and phase fluctuations, and yields:

$$S_+(k) = \int_{-\infty}^{\infty} \frac{T}{w^2 + (2p + k^2)^2} dw = \frac{T\pi}{k^2 + 2p} \tag{5.a}$$

$$S_-(k) = \int_{-\infty}^{\infty} \frac{T}{w^2 + k^4} dw = \frac{T\pi}{k^2}. \tag{5.b}$$

This means that the spectrum of phase fluctuations is of the form $1/k^2$ (5.b). The spatial spectrum of amplitude fluctuations is Lorentzian: in the short wavelength limit, $|k|^2 \gg 2p$, the amplitude spectrum is equal to the phase spectrum $S_+(k) = S_-(k)$. The total spectrum is: $S(k) = 2S_+(k)$ in this short wavelength limit. In the long wavelength limit $|k|^2 \ll 2p$, the amplitude fluctuation power spectrum saturates to: $S_+(k \approx 0) = T\boldsymbol{p}/(2\boldsymbol{p})$, and is negligibly small compared to the phase fluctuation spectrum. Therefore the total spectrum is essentially determined by the phase fluctuations in this long wavelength limit.

**b) temporal power spectra**

The temporal power spectra are obtained by integration of (4) over all possible spatial wavevectors **k**. In the case of one spatial dimension:

$$S_{+1D}(\boldsymbol{w}) = \int_{-\infty}^{\infty} \frac{T}{\boldsymbol{w}^2 + (2\boldsymbol{p} + k^2)^2} dk = \frac{T\boldsymbol{p}}{\boldsymbol{w}} \text{Im}\left[(2\boldsymbol{p} - i\boldsymbol{w})^{-1/2}\right] \quad (6.a)$$

$$S_{-1D}(\boldsymbol{w}) = \int_{-\infty}^{\infty} \frac{T}{\boldsymbol{w}^2 + k^4} dk = \frac{T\boldsymbol{p}}{2^{1/2} \boldsymbol{w}^{3/2}}. \quad (6.b)$$

This results in a power spectrum of phase fluctuations (6.b) of precisely the form of $1/\boldsymbol{w}^{3/2}$ in the entire frequency range (6.b). The spectrum of amplitude fluctuations (6.a) is more complicated: in the limit of large frequencies $|\boldsymbol{w}| \gg 2\boldsymbol{p}$ it is equal to the phase spectrum $S_{+1D}(\boldsymbol{w}) = S_{-1D}(\boldsymbol{w})$. In the limit of small frequencies $|\boldsymbol{w}| \ll 2\boldsymbol{p}$, the amplitude power spectrum saturates to: $S_{+1D}(\boldsymbol{w} \approx 0) = T\boldsymbol{p}/(2 \cdot (2\boldsymbol{p})^{3/2})$. In this way (6.a) represents a Lorenz-like spectrum (with a $1/\boldsymbol{w}^{3/2}$ asymptotic frequency dependence instead of $1/\boldsymbol{w}^2$ asymptotic frequency dependence of Lorentzian spectrum) for the amplitude fluctuations for systems extended in one-dimensional space.

The integration of (4) in two spatial dimensions yields:

$$S_{+2D}(\boldsymbol{w}) = \frac{T\boldsymbol{p}}{2\boldsymbol{w}}(\boldsymbol{p} - 2 \cdot arctg(2\boldsymbol{p}/\boldsymbol{w})), \qquad S_{-2D}(\boldsymbol{w}) = \frac{T\boldsymbol{p}^2}{2\boldsymbol{w}} \quad (7)$$

This results in a power spectrum of phase fluctuations in the form $1/\boldsymbol{w}$ in the entire frequency range, and in a Lorenz-like spectrum of amplitude fluctuations with exponent $\boldsymbol{a} = 1$. Finally, the integration of (4) in 3D yields:

$$S_{+3D}(\boldsymbol{w}) = \frac{2T\boldsymbol{p}^2}{\boldsymbol{w}} \text{Im}\left[(2\boldsymbol{p} + i\boldsymbol{w})^{1/2}\right], \qquad S_{-3D}(\boldsymbol{w}) = \frac{T\boldsymbol{p}^2 2^{1/2}}{\boldsymbol{w}^{1/2}} \quad (8)$$

This results in an exponent $\boldsymbol{a} = 1/2$ of amplitude and phase power spectra.

Generalizing, the power spectra of phase fluctuations for D-dimensional systems (e.g. for fractal dimensional systems) are $S_{-D}(f) \approx 1/f^{\boldsymbol{a}}$ with $\boldsymbol{a} = 2 - D/2$. For the amplitude fluctuations one obtains Lorenz-like power spectra, saturating for small frequencies, and with $1/f^{\boldsymbol{a}}$ dependence for high frequencies with the same exponent $\boldsymbol{a}$. The width of the Lorenz-like power spectra of amplitude fluctuations depends on the supercriticality parameter $p$: $\boldsymbol{w}_0 \approx 2|p|$.

## 4. Numerical results

The spectral densities (5)-(8) calculated from the linearisation were compared to those obtained directly by numerical integration of CGLE (1) in 1, 2, and 3 spatial dimensions. A split-step numerical integration technique was used calculating the local terms (linear gain, nonlinearity) in the spatial domain and nonlocal terms (Laplace operator) in spatial Fourier domain. Calculations were performed using periodic boundary conditions on a spatial grid of (32*32*32) in the 3D case, (128*128) in the 2D case, and ($2^{13}$) in the 1D case. The CGLE with real-valued coefficients $b = c = 0$ was numerically integrated with supercriticality parameter $p = 1$.

### a) temporal power spectra

The numerically calculated temporal power spectra are plotted in Fig.1. The $1/f^a$ character of the noise spectra is most clearly seen in the case of a 1D system (here $a = 3/2$). In two dimensions the $1/f^a$ noise ($a = 1$) is visible over almost three decades of frequency, and in three dimensions ($a = 1/2$) over almost two decades. The dashed lines in Fig.1 indicate the expected slopes.

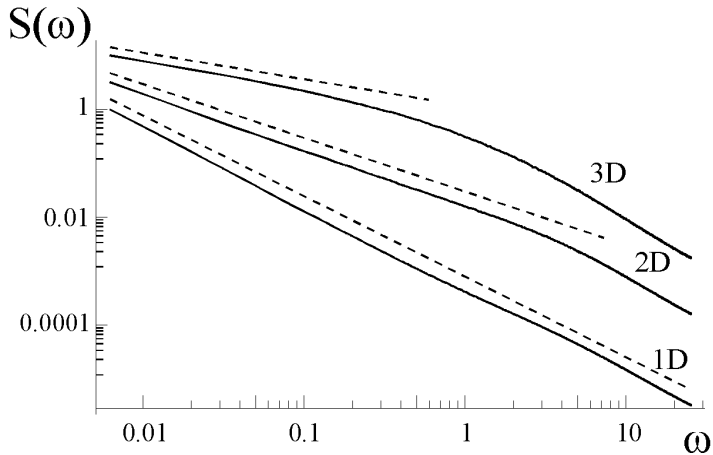

**Fig.1.** Total temporal noise power spectra in 1, 2 and 3 spatial dimensions, as obtained by numerical integration of the CGLE. Dashed lines show the slopes $a = 1/2$, $a = 1$, and $a = 3/2$. The spectra are arbitrary displaced vertically to distinguish between them. Integration period is 1000; the averaging was performed over 2500 realizations.

The main obstacle to calculate numerically the noise spectra in the entire frequency range is the discretization of the spatial coordinates and of time in the integration scheme. Discretization of space imposes a truncation of the higher spatial wavenumbers, and thus most strongly affects the high frequency components of the temporal spectra. Therefore to obtain numerically the spectra over the entire frequency range a series of separate calculations for different integration regions were performed and the spectra in the corresponding frequency ranges were combined them into one plot. (The size of integration region $L$ imposes the lower truncation bound of wavenumbers $k_{min} \approx 2p/L$, and the size of integration grid $l$ imposes the upper truncation bound of wavenumbers $k_{max} \approx 2p/l$. Roughly speaking the temporal spectra are then correct in the frequency range $w_{min} < w < w_{max}$, $w_{min,max} \approx k^2_{min,max}$ as (3,4) hint.) The calculations shown in Fig.2 were performed for the 2D case with four different sizes of integration regions $l = l_n = 2p \cdot 10^{2.5-n/2}$ ($n = 1,...,4$). The spectrum combined from partially overlapping pieces results in a 1/f dependence extending over more than five decades in frequency. The "kink" separating the low frequency range (with the amplitude fluctuations negligible compared to the phase fluctuations) and high frequency range (with the amplitude fluctuations equal to the phase fluctuations) is visible in the log-log represented power spectrum in Fig.2.a, and especially in the normalized power spectrum $wS(w)$ in Fig.2.b.

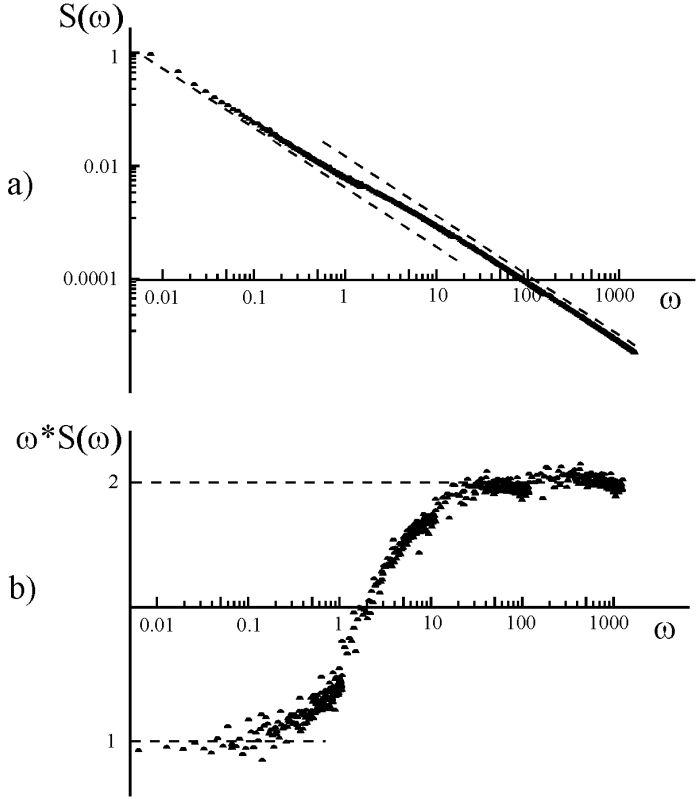

**Fig.2.** Total temporal noise power spectra in 2 spatial dimensions, as obtained by numerical integration of the CGLE. Integration period is $10^7$ temporal steps; the averaging was performed over 2500 realizations. The calculations were performed with 4 different sizes of the integration region with different temporal steps:
$l_1 = 2p \cdot 10^2$, $\Delta t_1 = 5 \cdot 10^{-2}$;
$l_2 = 2p \cdot 10^{1.5}$, $\Delta t_2 = 5 \cdot 10^{-3}$;
$l_3 = 2p \cdot 10$, $\Delta t_3 = 5 \cdot 10^{-4}$;
$l_4 = 2p \cdot 10^{0.5}$, $\Delta t_3 = 5 \cdot 10^{-5}$;
and these spectra were combined into one plot in log-log representation. a) power spectra. The dashed lines correspond to $1/w$ and are for guiding the eye; b) normalized power spectra $wS(w)$. The normalized power spectra in the limit of low- and high frequencies differ by a factor of 2.

A multiscale numerical integration of CGLE in 1D and 3D was also performed. It showed the $1/f^{3/2}$ - and $1/f^{1/2}$ - dependencies respectively over more than 5 decades of frequency (not shown).

**b) spatial power spectra**

Numerical discretization also distorts the spatial spectra, since it restricts the range of spatial wavenumbers. Therefore we performed a series of calculations with different sizes of the integration region, and combined the calculated averaged spatial spectra into one plot. The calculations in Fig.3 (2D case) were performed with five different sizes of the integrating region $l = l_n = 2p \cdot 10^{2.5-n/2}$ ($n = 1,...,5$). In this way we obtained the spectra combined from partially overlapping pieces, extending in total over around four decades.

Fig.3.a show the spectra in a log-log representation, where the character of $1/k^2$ is clearly seen, especially in the limits of long and short wavelengths. The "kink" at intermediate values of *k*, as seen most clearly seen from Fig.3.b, joins the spectra in the limits of long and short wavelengths which are both of the same slope, but of different intensities.

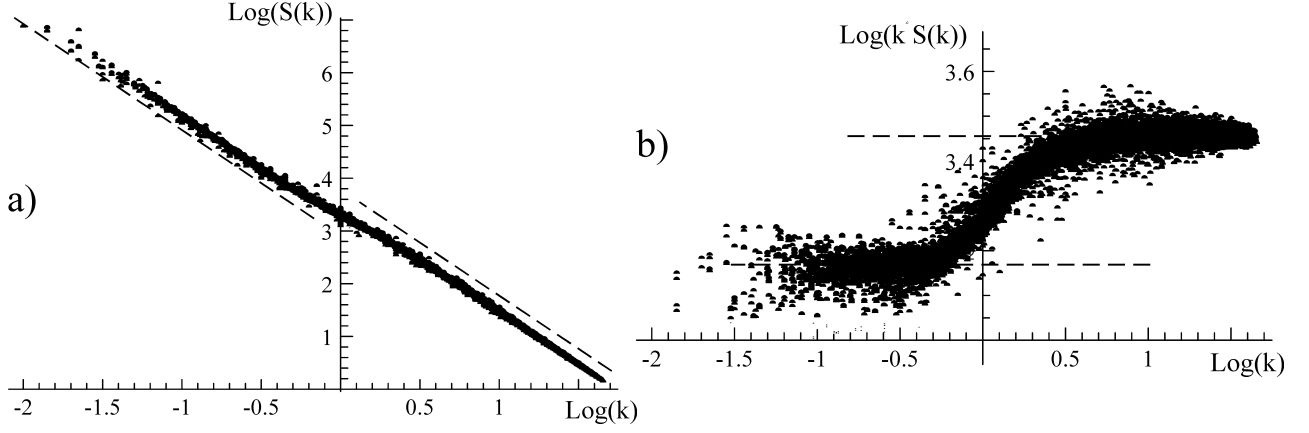

**Fig.3.** Total spatial noise power spectra in 2 spatial dimensions, as obtained by numerical integration of the CGLE. The averaging was performed over the time of $10^7$ temporal steps. Each point corresponds to the averaged intensity of a discrete spatial mode. The calculations have been performed with 5 different values of the size of the integration region with different temporal steps:

$l_1 = 2p \cdot 10^2$, $\Delta t_1 = 5 \cdot 10^{-2}$;

$l_2 = 2p \cdot 10^{1.5}$, $\Delta t_2 = 5 \cdot 10^{-3}$;

$l_3 = 2p \cdot 10$, $\Delta t_3 = 5 \cdot 10^{-4}$;

$l_4 = 2p \cdot 10^{0.5}$, $\Delta t_3 = 5 \cdot 10^{-5}$;

$l_5 = 2p$, $\Delta t_3 = 5 \cdot 10^{-6}$.

These spectra were combined into one plot log-log representation. a) power spectra. The dashed lines correspond to a $1/k^2$ dependence and are for guiding the eye; b) normalized power spectra $k^2 S(k)$. The normalized power spectra in the limit of low- and high spatial wavenumbers differ by a factor of 2.

One more reason to combine the spectra from pieces calculated separately was the finite size of the temporal step used in the split-step numerical technique. Indeed, in order to obtain the correct spatial spectra in the long wavelength limit the integration is time-consuming. The long waves are very slow, and the characteristic build-up time for long waves is of order of $t_{build} \approx 1/k^2$, as seen from (4,5), and diverges for $k \to 0$. Thus one has to average for very long time to obtain the correct statistics for the long waves. On the other hand, the characteristic build-up times for short wavelengths are very small, since the same relation $t_{build} \approx 1/k^2$ holds. Here, in order to obtain a correct statistics of mode occupation one has to correspondingly decrease the size of the temporal step for $k \to \infty$. We thus come to the conclusion, that one can never obtain the analytically predicted (correct) $1/k^2$ statistical distribution in a single numerical run, with finite temporal steps (with limited time resolution). The spectrum calculated with a fixed temporal step is shown in Fig.4. In log-log representation (Fig.4.a) a sharp decrease of occupation of the large wavenumbers occurs. In representation of logarithm of spectral density vs. $k^2$ (Fig.4.b) a straight line indicating an exponential decrease is obtained for large wavenumbers. The spectrum shown in Fig.4, curiously enough, is thus precisely a Bose-Einstein distribution, decaying with the power law for long wavelengths: $S(k \to 0) \propto k^{-2}$, and exponentially for short wavelengths: $S(k \to \infty) \propto \exp(-k^2)$.

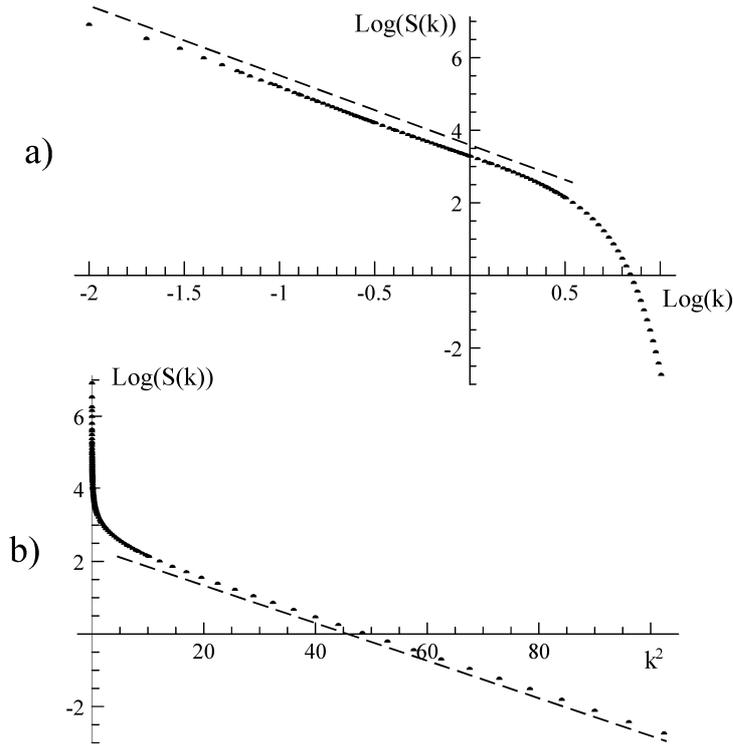

**Fig.4.** The total spatial spectrum as obtained by numerical integration of CGLE for 1D for fixed temporal step of $\Delta t = 5 \cdot 10^{-2}$, but combined from four calculations with different size of integration region. The averaging was performed over the time of $t = 10^6$. a) shows the spectrum in log-log representation and the dashed line corresponds to $1/k^2$ dependence; b) shows spectrum in single log representation and the dashed line corresponds to $\exp(-k^2)$ dependence

We note, that the linear stability analysis does not lead to the Bose-Einstein distribution as found numerically with finite temporal steps. The finite temporal step $\Delta t$ is equivalent to a particular cut-off frequency $w_{max}$ of the temporal spectrum: $w_{max} = 2p/\Delta t$. In order to account for this finite temporal resolution the integration of (5) should be performed not over the all frequencies, but over $[0, w_{max}]$. This integration, however, leads to a power law decay $S(k \to \infty) \propto k^{-4}$ for short wavenumbers, and not to the expected exponential decay. We have no explanation for this discrepancy between the analytical and numerical results.

We performed a series of numerical calculations varying the size of the temporal step, in order to interpolate the spectra in the total range of the spatial frequencies. The result is:

$$S(k) = \frac{TpC/w_{max}}{\exp(k^2 C/w_{max}) - 1}. \tag{7}$$

Here $C$ is a constant of order one. (7) reproduces correctly the numerically obtained spectra in both asymptotics of $k \to 0$ and $k \to \infty$. For the intermediate values of the wavelengths $k^2/w_{max} \approx 1$ a transition between power law and exponential decay is predicted by (7) exactly as it found in the numerical calculations. In this way numerical results show that the spatial spectrum of the CGLE in the case of limited temporal resolution coincide precisely with the Bose-Einstein distribution, whereas the spectrum in the case of unlimited temporal resolution follows the power law.

## 5. Conclusions

Concluding, we show analytically and numerically that the power spectra of spatially extended systems with order-disorder transitions obey power laws: The spatial noise spectra are of $1/k^2$ - form, thus being Bose-Einstein-like. The temporal noise spectra of the CGLE are shown to be of $1/f^a$ - form, with exponent $a = 2 - D/2$ depending explicitly only on the dimension of space *D*. Spatially extended systems with order-disorder transitions are described by a CGLE with stochastic forces (1), as the equation accounting for the symmetries of the phase space (Hopf bifurcation) and the symmetries of the physical space (rotational and translational invariance).

All ordered states in nature are presumably 1 to 3 - dimensional. This correspond to the exponents of $1/f^a$ noise of $1/2 < a < 3/2$, according to our model, which corresponds well to experimentally observed exponents of $1/f^a$ noise. The exponent *a* is found experimentally in the range $0.6 \leq a \leq 1.4$ [P.Dutta & P.M.Horn, 1981; Sh.M.Kogan, 1985, M.B.Weissman, 1985; G.P.Zhigal'skii, 1997] depending on the particular system. Another prominent feature of 1/f noise is that the spectra usually extend over many decades of frequency with constant *a*, which also follow simply and naturally from our model.

Our model for 1/f noise comprises the two most accepted models for 1/f noise. In [F.K.du Pre, 1950; A.Van der Tiel, 1950] 1/f noise is interpreted as a result of the superposition of Lorentzian spectra requiring a somewhat unphysical assumption of a specific distribution of damping rates. In our model, formally the 1/f spectra also result from a superposition of stochastic spatial modes (3) and (4). However, the distribution of the damping rates $f(g)$ ($g = \mathbf{k}^2$ in our case) results naturally from the dimensionality of the space and is universally valid.

There is also a relation with the model of self-organized criticality [P.Bak, C.Tang & K.Wiesenfeld, 1987; [P.Bak, C.Tang & K.Wiesenfeld, 1988], in that the phase variable in our model is always in a critical state, as (2.b) indicates. This analogy to self-organized criticality for the phase variable is a consequence of the phase invariance in the Hopf bifurcation. Consequently, one would expect that the noise power spectra of the models of self-organized criticality would show the same dependence on the spatial dimension $a = 2 - D/2$ as found here. To our knowledge no detailed investigations of the dependence of *a* on the dimension of space has been performed to date for self-organized criticality.

The above dependence of *a* on dimension of space leads to general conclusions concerning the stability of the ordered state of the system. The integral of the $1/f^a$ power spectrum always diverges in the limit of large or of small frequency, indicating a breakup of the ordered state in the limit of small or of large times, respectively. E.g. in case of low dimensional systems: $D < 2$, $a > 1$ the integral of temporal power spectra diverges at low frequencies, which means that the average size of fluctuation of order parameter grows to infinity for large times. The average size of fluctuation is: $\langle |a(t)|^2 \rangle \approx \int_{w_{\min}}^{\infty} S(w)dw$, where $w_{\min} \approx 2p/t$ is the lower cut-off boundary of temporal spectra, thus grows as $\langle |a(t)|^2 \rangle \propto t^{a-1}$ with increasing time. This generalizes the well known Wiener stochastic diffusion process $\langle |a(t)|^2 \rangle = t$ well known for zero dimensional systems, and predicts, that the diffusion in spatially extended systems is weaker than in zero dimensional systems. E.g. the fluctuations of the order parameter in 1D systems ($a = 1.5$) should diffuse as $\langle |a(t)|^2 \rangle = t^{1/2}$. This also means that for large times the fluctuations of the order parameter become on average of the order of magnitude of the order parameter itself. Defects must then appear in the ordered state, even for a small temperature.

For high dimensional systems $D > 2$, $a < 1$, contrarily, the integrals over temporal power spectra diverge at large frequencies. It may be expected that large fluctuations of order parameter occur at small times: $\langle |a(t)|^2 \rangle \approx \int_0^{w_{max}} S(w)dw$, where $w_{max} \approx 2p/t$ is the upper cut-off boundary of temporal spectra. This results in $\langle |a(t)|^2 \rangle \propto 1/t^{1-a}$ diffusion law for $a < 1$, diverging for small times. E.g. the fluctuations of the order parameter in 3D systems ($a = 0.5$) should diverge as $\langle |a(t)|^2 \rangle = t^{-1/2}$ for small times. This means that a continuos creation and annihilation of pairs of defects of the ordered states for $D > 2$ can be predicted. (These defects are termed "virtual defects", since they appear on a short time scale only and do not bear dynamical significance.)

The case $D = 2$ is marginal. The integral over the spectrum diverges weakly (logarithmically) in both limits of small and of large frequencies. Specifically, such low frequency divergence in 2D patterns was investigated in [J.M.Kosterlitz & D.J.Thouless, 1973], and is termed Kosterlitz-Thouless transition.

The work has been supported by Sonderforschugsbereich 407 of Deutsche Forschungsgemeinschaft. Discussions with C.O.Weiss, M.Lewenstein, and A.Berzanskis are acknowledged.